\documentclass[superscriptaddress,aps,prl,twocolumn,preprintnumbers,amsmath,amssymb,floatfix,groupedaddress]{revtex4-1}
\usepackage[dvips]{graphicx}
\usepackage{dcolumn}
\usepackage{bm}
\usepackage{xr}
\usepackage[driverfallback=dvipdfm]{hyperref}
\usepackage{color}

\begin{document}

\title{Homodyne tomography of a single photon retrieved on demand from a cavity-enhanced cold atom memory}
\author{Erwan Bimbard}
\email{erwan.bimbard@institutoptique.fr}
\author{Rajiv Boddeda}
\author{Nicolas Vitrant}
\author{Andrey Grankin}
\author{Valentina Parigi }
\altaffiliation[Present address: ]{Laboratoire Kastler Brossel, Universit\'e Pierre et Marie Curie, \'Ecole Normale Sup\'erieure, CNRS, 4 place Jussieu, 75252 Paris Cedex 05, France}
\author{Jovica Stanojevic}
\author{Alexei Ourjoumtsev}
\author{Philippe Grangier}
\affiliation{Laboratoire Charles Fabry, Institut d'Optique, CNRS, Univ.
Paris Sud, 2 av. Augustin Fresnel, 91127 Palaiseau cedex, France }

\begin{abstract}
We experimentally demonstrate that a non-classical state prepared in an atomic memory can be efficiently transferred to a single mode of free-propagating light. By retrieving on demand a single excitation from a cold atomic gas, we realize an efficient source of single photons prepared in a pure, fully controlled quantum state. We characterize this source using two detection methods, one based on photon-counting analysis, and the second using homodyne tomography to reconstruct the density matrix and Wigner function of the state. The latter technique allows us to completely determine the mode of the retrieved photon in its fine phase and amplitude details, and demonstrate its non-classical field statistics by observing a negative Wigner function. We measure a photon retrieval efficiency up to $82 \%$ and an atomic memory coherence time of $900$ ns. This setup is very well suited to study interactions between atomic excitations, and to use them in order to create and manipulate more sophisticated quantum states of light with a high degree of experimental control.
\end{abstract}

\maketitle


Precisely controlling quantum states of the light is crucial for many quantum information processing (QIP) tasks. In this respect, cold atomic ensembles are a very versatile tool. In the past decade, they have been extensively used to store and retrieve quantum states of light \cite{Lvovsky2009b} and, more recently, to manipulate them using atomic interactions \cite{Dudin2012,Peyronel2012,Maxwell2013}. Quite often, however, the resulting quantum states were characterized only partially, typically by photon correlation measurements insensitive to the exact frequency or temporal envelope of the emitted photons. Although independent atomic ensembles have been shown to emit photons with a good degree of indistinguishability \cite{Thompson2006,Felinto2007,Chaneliere2007,Yuan2007}, their exact wavefunctions remained to be characterized. Demonstrating that strongly non-classical states stored in an atomic ensemble can be retrieved on demand as single-mode, Fourier-transform limited light pulses is an important requirement to use such systems in QIP and to connect them to other QIP devices.

A very convenient way to fully characterize the retrieved quantum state is to use homodyne tomography \cite{Lvovsky2009}, by reconstructing the photons' density matrix and Wigner function from the quadrature distributions of the light field measured by interference with a bright local oscillator. Since this technique characterizes the light's field rather than its intensity, it is intrinsically single-mode and very sensitive to optical losses. Therefore, observing non-classical features in the reconstructed quantum state proves that it can truly be emitted on demand, detected with a high efficiency, and controlled in all of its degrees of freedom which must exactly match those of the local oscillator. For a non-classical state, the field's quadrature statistics are described by a non-Gaussian Wigner function taking negative values: losses or imperfections in the spatial mode, the polarization, the temporal envelope or the frequency of the state will degrade the signal and eventually make its Wigner function completely positive. This makes homodyne characterization more precise but also technically more challenging than photon correlation measurements, which can be made insensitive to many of these imperfections. Ubiquitous in the characterization of Gaussian continuous-variable quantum memories \cite{Honda2008,Appel2008,Cviklinski2008,Jensen2011,Hosseini2011}, this technique has been extended during the last decade to analyse non-Gaussian states produced with heralded parametric or hot atomic vapor light sources \cite{Lvovsky2009,MacRae2012}, and very recently with an optical quantum memory \cite{Yoshikawa2013}. Transposing it to a cold atomic ensemble, where quantum light states can not only be stored but also processed, should form an ideal system for studying non-linear transformations of few-photon states due to atomic interactions.

In this letter we describe an experimental homodyne tomography of a single photon retrieved on demand from a cloud of cold Rubidium atoms placed within an optical cavity. We show that both our atomic system and our detection setup are sufficiently performant to preserve the negativity of the Wigner function directly reconstructed from raw experimental data. Taking into account the imperfections of the homodyne detector allows us to characterize the atomic system alone and to show that it can operate as a quantum memory with a 900 ns storage time, emitting free-propagating single photons in a well-controlled single spatiotemporal mode with an efficiency up to $82\%$.

\begin{figure}[ht]
\centerline{
\includegraphics[scale = 0.5]{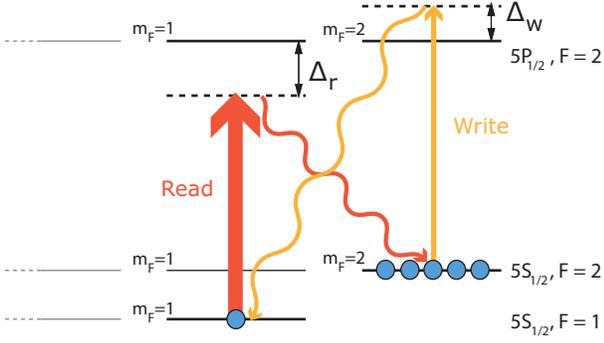}
}
\caption{DLCZ protocol. In a cloud of $^{87}$Rb atoms, two Raman transitions are used to store (``Write'') and retrieve (``Read'') a single photon as a collective excitation. Series of consecutive pulses of Write and Read beams are sent, with detunings $\Delta_W$ and $\Delta_R$ respectively.
\label{fig:dlcz}}
\end{figure}


Our single-photon source is based on the DLCZ protocol proposed by Duan et al \cite{Duan2001}, implemented in a cloud of cold $^{87}$Rb atoms (Fig. \ref{fig:dlcz}), in a pulsed scheme. In the weak driving limit, detecting a spontaneous Raman ``Write'' photon heralds the creation of a single collective atomic spin excitation, which can be stored for a variable time and retrieved as a single ``Read'' photon by another Raman step. The collective, phase-matched character of the four-photon process greatly enhances the photon retrieval efficiency in the observed mode, further improved by placing the atoms in a low-finesse optical cavity resonant with the emitted photons \cite{Simon2007}. The maximal theoretical retrieval efficiency $\eta_{max}=C/(1+C)$ depends only on the cooperativity $C=Ng^2/\kappa \gamma$ of the $N$ atoms coupled to the cavity field with a single-atom vacuum Rabi frequency $g$, $\gamma$ and $\kappa$ being respectively the atomic dipole and the cavity field decay rates. In order to retrieve a Fourier-limited single-photon pulse with a constant phase profile, the frequency and time dependence of the Read laser pulse are adjusted according to Ref. \cite{Stanojevic2011}. In practice, the retrieval efficiency is also reduced by various defects and decoherence sources.

\begin{figure}[ht]
\centerline{
\includegraphics[scale = 0.45]{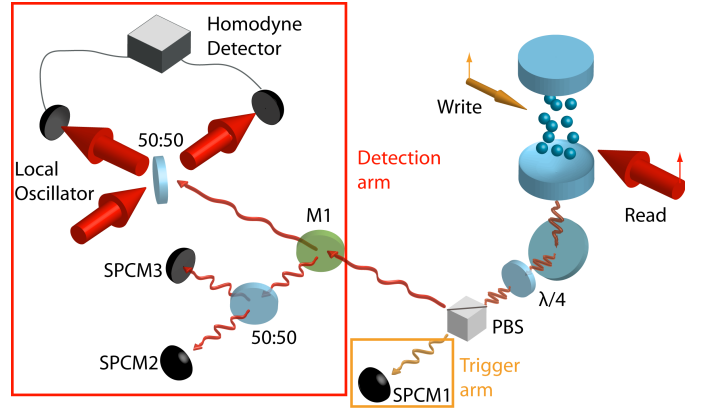}
}
\caption{Experimental setup. A cloud of cold Rubidium atoms trapped inside an optical cavity is off-resonantly excited with alternating Write and Read pulses, creating pairs of photons Raman-scattered in the cavity. They are split according to their polarization on a Polarizing BeamSplitter (PBS). The detection of a Write photon by the Single Photon Counting Module SPCM 1 heralds the creation of a single atomic spin excitation, retrieved by the read pulse as a single photon directed either to SPCMs 2 and 3 for photon-counting measurements, or to a homodyne detection setup. Mirror M1 is flipped in or out according to the choice of detector.
\label{fig:setup}}
\end{figure}

The experimental setup is sketched on figure \ref{fig:setup}. A magneto-optically trapped cloud of $^{87}$Rb is polarization-gradient cooled to $\approx 50\ \mu$K and released in free fall inside a 66-mm long vertical Fabry-Perot cavity with a finesse $F = 120$ and a linewidth $\kappa/2\pi = 10$ MHz. The cavity sustains a fundamental mode of waist $w_0 = 86\ \mu$m at $795$ nm, coupled to $\sim 10^4$ atoms, providing the system with a cooperativity $C \approx 15$ on the D1 transition. The difference in the cavity mirrors' reflectivities ( $R^{top}=99.99 \%$, $R^{bottom}=95 \%$) ensures that the photons emitted in this mode escape the cavity through the bottom mirror with a nearly $100\%$ probability.

After an initial $100\ \mu s$ step of optical pumping to the state $5S_{1/2},F=2,m_F=+2$, the atoms are excited with a series of counter-propagating Write and Read laser pulses, alternating with a $1.5$ $\mu s$ period during $1$ ms, before optical pumping quality degrades. This pair of pumping/measurement steps is then repeated 12 times before the atoms are recaptured and recooled for $120$ ms and the sequence starts again, allowing us to perform $\approx 60 000$ Write/Read trials per second. The Write and Read pulses are both $\pi$-polarized and propagate orthogonally to the quantization axis, oriented along the cavity. Their temporal intensity profiles are nearly Gaussian, with a $1/e$ half-width $\sigma_w = \sigma_r = 80$ ns and an adjustable delay. The Write pulse is detuned by $\Delta_W/2\pi = +28$ MHz from the $5S_{1/2},F=2,m_F=2\rightarrow 5P_{1/2},F=2,m_F=2$ transition, making the Write photon emitted by Raman scattering towards the $5S_{1/2},F=1,m_F=1$ state resonant with a TEM$_{00}$ cavity mode. Similarly, after accounting for a linear resonance shift induced by the atoms, a Read pulse detuning of $\Delta_R/2\pi = -46$ MHz from the $5S_{1/2},F=1,m_F=1\rightarrow 5P_{1/2},F=1,m_F=1$ makes the emitted Read Raman photon bringing the system back to the initial $5S_{1/2},F=2,m_F=+2$ state resonant with another TEM$_{00}$ cavity mode, shifted from the previous one by 3 cavity free spectral ranges. After escaping the cavity, the photons are separated according to their opposite circular polarizations. The Write photons are then spatially and spectrally filtered via a single-mode optical fiber, interference filters and a Fabry-Perot cavity, designed to block stray light from other beams, before being detected with a Single Photon Counting Module SPCM 1. With a maximal Rabi frequency $\Omega_W^{max} / 2\pi \approx 200$ kHz for the Write pulse, and taking into account the $33\%$ total Trigger path detection efficiency, a Write photon is detected on SPCM1 with a probability $\approx 10^{-3}$ per Write pulse, which makes the probability to create multiple excitations or to depump the atoms negligible while keeping the trigger rate sufficiently high. On the other hand, the large maximal Read Rabi frequency $\Omega_R^{max} / 2\pi \approx 20$ MHz ensures that Read photons are retrieved with a high efficiency. After leaving the cavity these photons are directed either to a photon-correlation or to a homodyne detector, providing complementary information about the measured quantum state. Experimental control is handled by fast FPGA logic which allows us to acquire their signals only when a trigger photon has been detected during the previous Write pulse.

\begin{figure}[ht]
\centerline{
\includegraphics[scale = 0.55]{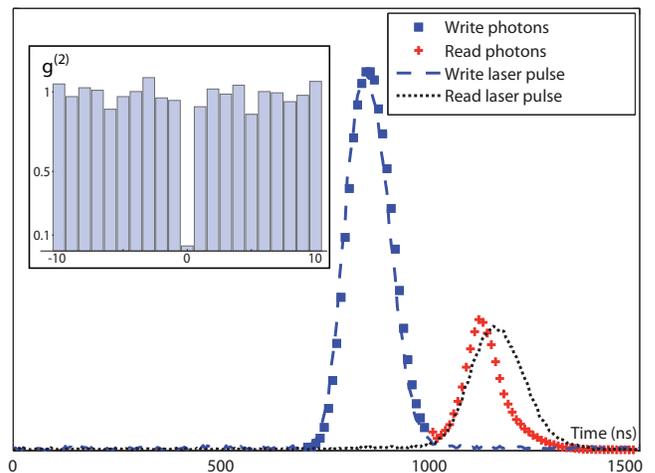}
}
\caption{Photon-counting measurements. Histograms of arrival times of the Write photon on SPCM1 and of the Read photon on SPCM 2 or 3, superimposed with the measured pulse profiles of the Write and Read pulses (arbitrary vertical units). Inset : measured correlation function $g^{(2)}$ showing anti-correlation of the photons in a single Read pulse down to $g^{(2)}(0) \approx 0.04$.
\label{fig:counting}}
\end{figure}

The photon-counting measurement is realized by splitting the photon between two SPCMs 2 and 3, to characterize the single-photon source via the standard second-order intensity correlation function
\begin{equation}
g^{(2)}(\tau)=\frac{\langle \hat{a}^\dag(\tau)\hat{a}^\dag(0)\hat{a}(0)\hat{a}(\tau)\rangle}{\langle \hat{a}^\dag\hat{a}\rangle^2}= \frac{\langle n_2(0)n_3(\tau)\rangle}{\langle n_2\rangle\langle n_3\rangle}
\end{equation}
where $n_j$ is the number of photons measured by detector $j$ during a Read pulse. Since the data acquisition is triggered by Write photons detected at random times, in the expression of the average coincidence rate $\langle n_2(0)n_3(\tau)\rangle$ the parameter $\tau$ does not correspond to a physically meaningful time but to a number of trigger events on SPCM1. The result of this measurement, displayed on Fig. \ref{fig:counting}, shows that $g^{(2)}(\tau\neq 0)\approx 1$ meaning that the detection events are uncorrelated within different pulses, whereas within the same pulse $g^{(2)}(0)=0.04\pm 0.01$, confirming the strongly Sub-Poissonian character of the photon source.

In addition to this correlation function, the histogram of photon detection times registered by each detector with a $10$ ns resolution allows us to determine the temporal envelopes of Write and Read photons. As shown on Fig. \ref{fig:counting}, the Write photon emission profile exactly matches the temporal intensity dependence of the Write pulse, as expected for a weakly driven spontaneous Raman process. On the other hand, as predicted in \cite{Stanojevic2011}, the Read photon is extracted at the beginning of the Read laser pulse and is considerably shorter than the latter, remaining approximately Gaussian with a $1/e$ half-width of $40$ ns.

Finally, knowing the total efficiency $\eta_c=37 \pm 1.5\%$ of the photon counting detection system, we estimate the intrinsic efficiency of our photonic source, defined as the probability to create a free-propagating photon in the analyzed mode each time a trigger Write photon was detected, to $\eta = 79 \pm 3 \%$. This free-space retrieval efficiency, relevant for most applications, exceeds those observed in previous similar experiments where efficient polariton-to-photon conversions were undermined by outcoupling or filtering losses \cite{Simon2007,Bao2012}.

To analyse the quantum state of the retrieved photon by homodyne tomography, it is sent to interfere on a $50:50$ beamsplitter with a continuous-wave local oscillator (LO) beam. The lasers producing the Read and LO beams are phase locked, which allows us to precisely match the LO and the Read photon's frequencies. Approximately $7$ mW of LO power sets the vacuum noise level $\approx 20$ dB above electronic noise. The spatial mode-matching between the LO and the signal is optimized using a probe transmitted through the cavity locked on its fundamental mode. Since our state generation process is phase-independent, we do not actively control the phase between the Read and the LO beams and assume that the density matrix of the state is diagonal in the Fock basis (if this assumption was wrong, this would only lead to underestimating the state's purity).

\begin{figure}[ht]
\centerline{
\includegraphics[width = 8.6cm]{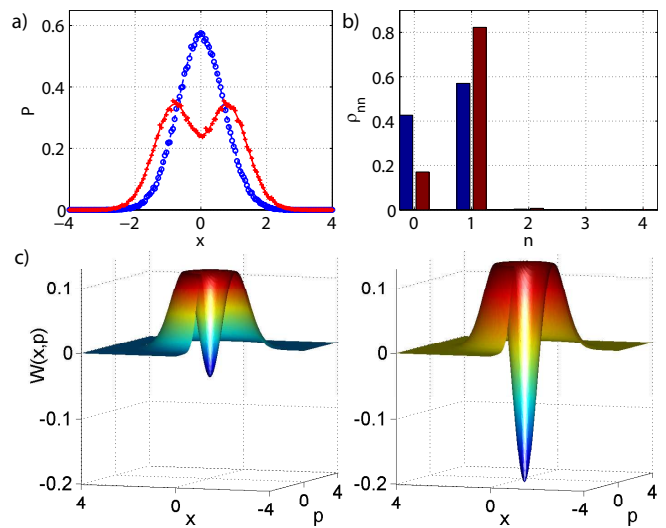}
}
\caption{Homodyne measurement of the single photon. a) Measured quadrature probability distribution $P_0(x)$ of the vacuum state  (blue circles) and $P_1(x)$ of the single photon  (red crosses), interpolated by the probability distributions estimated by integrating the reconstructed uncorrected Wigner functions along one dimension in phase space. b) Single-photon density matrix after maximum likelihood reconstruction before (left, blue) and after (right, red) correction for the detector's losses and noise. c) Corresponding Wigner functions.
\label{fig:homodyne}}
\end{figure}

Upon detection of a trigger photon, we register $2.2\ \mu$s of homodyne data sampled at a $250$ MHz rate. Each data sample $h_j(t)$ is then multiplied by a weighting function $f(t)$ and integrated. The obtained value $x_j=\int h_j(t)f(t)$ corresponds to the measurement of a field's quadrature $\hat{x}$ in the temporal mode defined by $f$. Since the homodyne measurement is sensitive to the field's amplitude, proportional to the square root of its intensity, the optimal weighting function is expected to be a Gaussian $\sqrt{2}$ times wider than the one measured by photon counting. This is confirmed by an independent optimization procedure which yields a $55\approx 40\times\sqrt{2}$ ns $1/e$ half-width for the optimal Gaussian weighting function. In order to properly normalize the measured signal, we measure the vacuum noise variance applying to the data the same filtering function $f$ shifted by $T\gg 55 ns$, thus selecting a temporally orthogonal mode containing a vacuum state. The measured quadrature probability distribution $P_0(x)$ of the vacuum state, shown on Fig. \ref{fig:homodyne}, is Gaussian, with a variance normalized to $1/2$ by convention. On the other hand, the quadrature distribution  $P_1(x)$ of the single photon, reconstructed by measuring $\approx 103 000$ quadrature values, is strongly non-Gaussian, with a characteristic dip around $x=0$. By applying a maximum likelihood algorithm to this quadrature distribution, we directly obtain the density matrix and the Wigner function of the measured state, displayed on Fig. \ref{fig:homodyne}. The density matrix presents a very small two-photon component and a single-photon fraction of $57\%$. As a result, the Wigner function reconstructed from raw data becomes negative at the origin. The marginal distributions obtained by integrating the vacuum and single-photon Wigner functions overlap very well with the measured quadrature distributions, confirming the consistency of the data analysis process.  The fact that the spectral width of the photon, obtained by Fourier-transforming its envelope, is narrower than those of the cavity and of the atomic transition (respectively $2$ MHz, $10$ MHz and $3$ MHz) shows that the emission process is not spontaneous but coherently driven.

In order to properly characterize the state produced by our source, we can integrate the imperfections of the homodyne detection setup in our maximum likelihood reconstruction procedure. This includes the optical transmission  between the vacuum chamber and the detector $\eta_{hd} = 82 \pm 1 \%$, the interference visibility between the LO and the signal $\eta_m = 96.5 \pm 1 \%$, the quantum efficiency of the detector's photodiodes $\eta_{q} = 91 \%$ and a $1\%$ electronic noise contribution. After accounting for these imperfections, the two-photon contribution to the density matrix remains low while the single-photon component increases to $\eta=82\pm 2\%$, in excellent agreement with the photon counting measurement, yielding a strongly non-classical negative Wigner function.

\begin{figure}[ht]
\centerline{
\includegraphics[scale = 0.45]{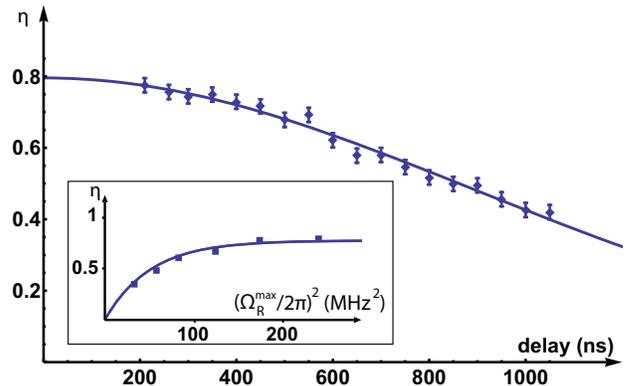}
}
\caption{Measured single photon retrieval efficiency $\eta$ as a function of the delay between the Write and Read pulses, fitted by a Doppler-induced Gaussian decay function with a characteristic time of $\approx 900$ ns. Inset : short-time retrieval efficiency as a function of the squared maximum Read Rabi frequency $(\Omega_R^{max})^2$, superimposed with a realistic model of the experiment (see text).
\label{fig:delay}}
\end{figure}

We have used a theoretical model of the experiment, based on the calculation of \cite{Stanojevic2011}, incorporating various defects, among which most importantly the imperfect optical pumping (evaluated to be $92\%$) and the spatial variations of the Write/Read laser beams and photon modes accross the atomic cloud. The maximal retrieval efficiency is consistent with the combination of the limit set by the finite cooperativity with these various imperfections in our setup. The optimal value $C\approx 15$ of the cooperativity, providing the best photon retrieval efficiency given the experimental imperfections, is very similar to the one obtained in previous experiments \cite{Simon2007}. The effect of a finite atomic temperature, inducing motional decoherence, can be measured by varying the delay between the Write and the Read pulses. As shown on Fig. \ref{fig:delay}, the time dependence of the retrieval efficiency follows the expected Gaussian decay. The measured characteristic time $\tau \approx 900$ ns is in very good agreement with the Doppler time for a Maxwell-Boltzmann distribution of atoms at $T = 50\ \mu$K ($\tau_D = \sqrt{m/k_B T}/2 k$, where $k$ is the wave vector of $795$ nm light, and $m$ the atomic mass of $^{87}$Rb). This Doppler decay could in principle be reduced by placing the atoms in a light-shift-compensated optical lattice \cite{Radnaev2010} or by increasing the wavelength of the atomic spin wave using an appropriate beam configuration \cite{Bao2012}, which should not fundamentally impact the purity of the retrieved photon. Our model also accounts for explicit dependence on the Read pulse profile and intensity. The agreement with the experiment is satisfactory, as shown for instance by the inset in Fig. \ref{fig:delay}, displaying the short-time retrieval efficiency as a function of the maximal Read beam Rabi frequency.

In summary, we experimentally demonstrate a cold-atom based single photon source that exhibits high modal purity and retrieval efficiency up to $82 \%$. A single excitation can be stored in the cloud with a coherence time of $900$ ns and retrieved on-demand as a single photon in a very well defined quantum state, consistently characterized by photon counting and homodyne tomography, revealing non-classical features in both the intensity and the light field. Since homodyne detectors can measure phase-dependent states with arbitrary photon numbers, this demonstrates that homodyne analysis methods can be conveniently applied to more complex state production schemes available in the cold atom ensemble systems, in particular those based on few-photon non-linearities induced by atomic interactions.

This work was supported by the EU through the ERC Advanced Grant 246669 ``DELPHI'' and the Collaborative Project  600645 ``SIQS''. We thank Marc-Henri Pottier for the assistance with the experimental control software.

%

\end{document}